\documentclass[aps,tightenlines,twocolumn,floatfix,showpacs]{revtex4}
\usepackage{graphics}
\usepackage{bm}
\usepackage{amsmath}
\usepackage{amssymb}

\begin{document}

\title{Nuclear mass systematics and nucleon-removal thresholds;
application to ${}^{17}$Na.}
  \author{K. Amos$^{1}$} 
  \email{amos@unimelb.edu.au}
  \author{D. van der Knijff$^{1}$}
  \author{L. Canton$^{2}$}
  \author{P. R. Fraser$^{3,1}$}
  \author{\mbox{S. Karataglidis$^{4}$}}
  \author{J. P. Svenne$^{5}$}

  \affiliation{$^{(1)}$ School  of Physics,  University of  Melbourne,
    Victoria 3010, Australia}
  \affiliation{$^{(2)}$ Istituto  Nazionale  di  Fisica  Nucleare,
    Sezione di Padova, Padova I-35131, Italia}
  \affiliation{$^{(3)}$ Instituto de Ciencias Nucleares,  Universidad
    Nacional Aut\'onoma de M\'exico, 04510 M\'exico, D.F., Mexico}
  \affiliation{$^{(4)}$ Department of Physics, University of Johannesburg, 
    P.O. Box 524 Auckland Park, 2006, South Africa}
  \affiliation{$^{(5)}$ Department  of  Physics  and Astronomy,
    University of Manitoba, and Winnipeg Institute for Theoretical Physics,
    Winnipeg, Manitoba, Canada R3T 2N2}

\date{\today}

\begin{abstract}
A survey of known threshold excitations of mirror systems
suggests a means to estimate masses of nuclear systems that are 
uncertain or not known, as does a trend in the relative energies 
of isobaric ground states.
Using both studies and known mirror-pair energy differences,
we estimate the mass of the nucleus ${}^{17}$Na and its energy
relative to the $p$+${}^{16}$Ne threshold. 
This  model-free estimate of the latter is larger than 
that suggested by recent structure models.
\end{abstract} 

\date{\today}

\maketitle

The spectra of radioactive nuclei are most intriguing,
especially of those at or just beyond a drip line. To date
not even the masses of many are known. Of
those for which some information does exist, details of spectra
are often poorly known at best. Few if any excited states
have been identified. Likewise, the spin-parities of many of
the states that are known have not been or are uncertainly assigned. 
However, the advent of radioactive ion beams and their scattering from
nuclei means that properties of quite exotic
nuclei can be sought. 
One  example is the specification of properties of the proton unstable
${}^{15}$F found from studies of ${}^{14}$O-$p$ scattering~\cite{Go04,Gu05}.

With light mass systems having charge number $\pi$ and neutron number
$\nu$, there is often the possibility to link the structures of mirror 
systems.  Usually there is  a reasonably well known spectrum of a nucleus
(${}_{\pi = Z}^{A+1}X_{\nu = N+1}$), which may be treated as a compound 
of a neutron ($n$) with ${}_{\pi = Z}^{\hspace*{0.4cm}A}X_{\nu = N}$ to 
define a nuclear model with which, assuming charge invariance of the 
nuclear force and adjusting for Coulomb effects, the spectrum of the 
mirror, ${}_{\pi = N+1}^{\hspace*{0.4cm}A+1}Q_{\nu = Z}$, may be predicted.

Of many past articles in which this symmetry has been used, just 
four~\cite{Fo05,Ca06a,Ti10,Fo10} are cited as examples.
In~\cite{Fo05}, a symmetry was applied with a shell model scheme, 
adapted to give the low
lying structure of ${}^{15}$C (from $n$+${}^{14}$C), and
spectra of other mass-15 nuclei deduced. That included
a spectrum for the proton-unstable ${}^{15}$F. In~\cite{Ca06a},
the same systems were studied using the collective approach
of a multichannel algebraic scattering (MCAS)~\cite{Am03} method.
In that study, not only was a spectrum of resonance states of ${}^{15}$F
predicted, but so also were low energy differential cross sections
from the scattering of radioactive ${}^{14}$O ions from hydrogen;
cross sections found to be in good agreement with
experiment~\cite{Go04,Gu05}. Subsequently~\cite{Mu09}, resonances were
found in the predicted energy region.

Recently,  Timofeyuk and Descouvemont~\cite{Ti10} and Fortune, Lacaze, and 
Sherr~\cite{Fo10} used microscopic structure models to define
the spectrum of ${}^{17}$C, treated as $n$+${}^{16}$C.
Then, using the charge symmetry argument, they ascertained 
a spectrum for ${}^{17}$Na (treated as $p$+${}^{16}$Ne).  
Both studies expect there to be a set of resonances in 
the low excitation spectrum of the particle unstable ${}^{17}$Na
and that the ground state of ${}^{17}$Na would
be a broad resonance of spin-parity $\frac{3}{2}^+$. The centroid energy
(width) has been predicted as 2.4 (1.36) MeV~\cite{Ti10} and
2.7 (2.2) MeV~\cite{Fo10}.  
However there is very little actually known about the ${}^{17}$Na system.
An early tabulation of nuclear masses~\cite{Wa88} put the mass excess 
(from theory) for ${}^{17}$Na at 35.61, 35.81, and 35.84 MeV. 
Using 35.61 MeV for the mass excess~\cite{Ke66}, 
the one- and three-proton emission
thresholds are 4.3 and 5.7 MeV below the
ground state of ${}^{17}$Na~\cite{Ti93} for break-ups of $p$+$^{16}$Ne and
for ${}^{14}$O + $3p$ respectively. The $3p$ threshold lies below that
of the single-proton emission since ${}^{16}$Ne is particle
emissive.

Defining two mirror nuclei by
$X = {}_{(\pi = Z)}^{\hspace*{0.5cm} A}X_{(\nu = N)}$
and
$Y = {}_{(\pi = N)}^{\hspace*{0.5cm} A}Y_{(\nu = Z)}$ ,
let the energies of the nucleon plus nucleus thresholds be
\begin{gather}
Th(nX)= E(n{\rm +}X) - E_{\rm g.s.}\left[
{}_{(\pi = Z)}^{(A+1)}X_{(\nu = N+1)} ,
\right]
\notag\\
\intertext{and} 
Th(pY)= E(p{\rm +}Y) - E_{\rm g.s.}\left[
{}_{(\pi = Z+1)}^{\hspace*{0.3cm}(A+1)}W_{(\nu = N)}
\right] .
\notag
\end{gather}
$\Delta(Th)$ denotes the difference $Th(nX)-Th(pY)$.
These energies for the mass-13 systems are 
\begin{align}
{}^{13}{\rm C}-{}^{13}{\rm N}\; :&\;\;
Th(n{\rm C}) = 4.95\;\; ;\;\;
Th(p{\rm C}) = 1.94
\nonumber\\
&\Delta(Th) = 3.01
\\
{}^{13}{\rm B}-{}^{13}{\rm O}\; :&\;\;
Th(n{\rm B}) = 4.88\;\; ;\;\;
Th(p{\rm N}) = 1.51
\nonumber\\
&\Delta(Th) = 3.37 . 
\end{align}
The data values leading to the $Th(nX),\; Th(pY)$, and  $\Delta(Th)$ were 
taken from the Ame2003 compilation~\cite{Au03}, as are all that are
specified and used hereafter.  

We seek a model-free scheme to
estimate ground state energies of nuclei as yet unmeasured or which are poorly 
established.
The isospin of the core nucleus will be used as a label for these energies. 
For example, the values for the mirror 
pair, ${}^{13}_{\ 6}$C$_7-{}^{13}_{\ 7}$N$_6$, are formed from the single, 
isospin $T=0$, core nucleus, ${}^{12}_{\ 6}$C$_6$, while those for the 
mirror pair ${}^{13}_{\ 5}$B$_8-{}^{13}_{\ 8}$O$_5$ have mirror
core nuclei with $T = 1$, ${}^{12}_{\ 5}B_7$ and ${}^{12}_{\ 7}N_5$
respectively.

\begin{figure}[h]
\scalebox{0.45}{\includegraphics*{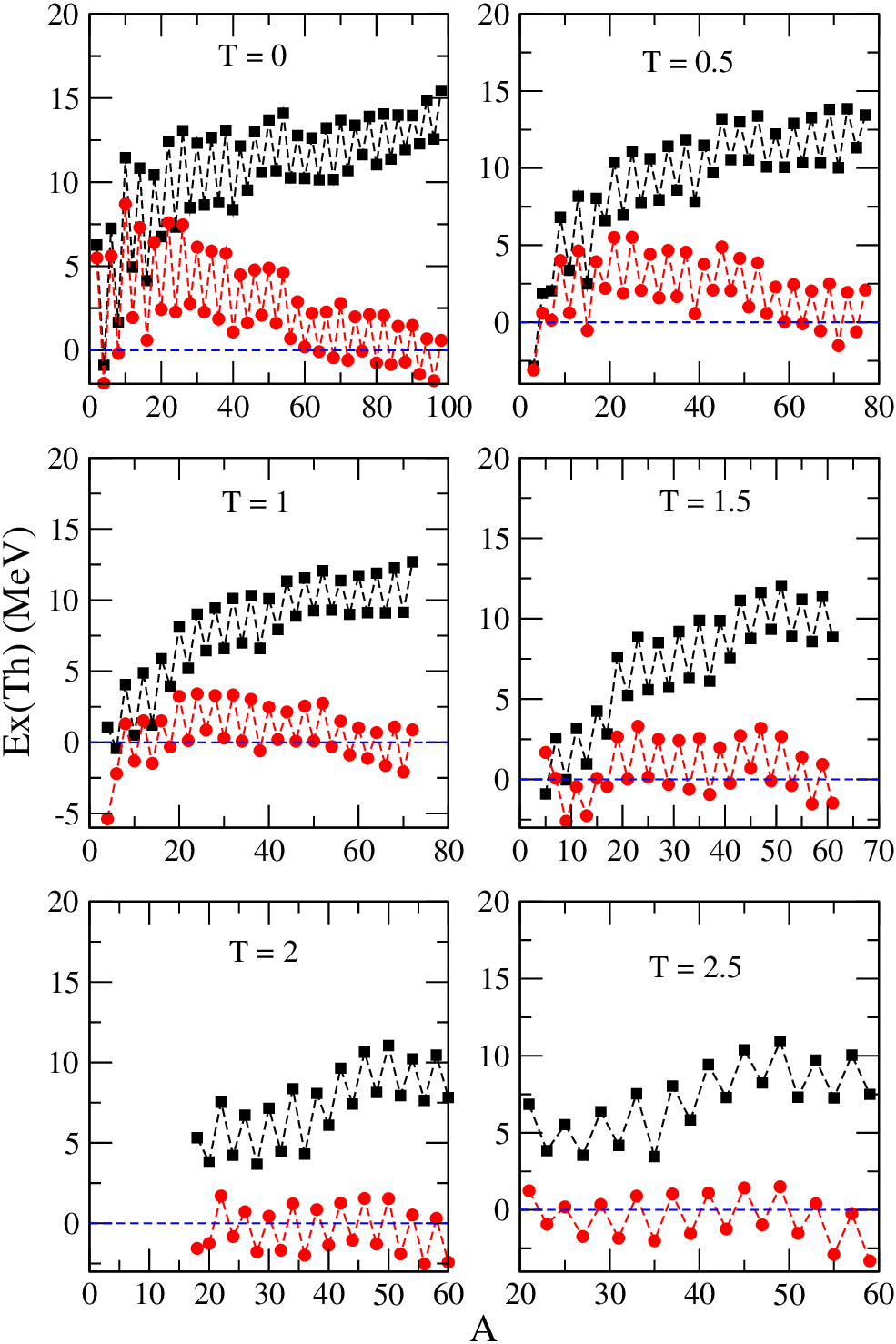}}
\caption{\label{Thresholds} (color online)
Excitation energies of particle-emission thresholds, 
$Th(nX)$ (filled squares) and $Th(pY)$ (filled circles).
The connecting lines are simply to guide the eye.}
\end{figure}
\begin{figure}[h]
\scalebox{0.45}{\includegraphics*{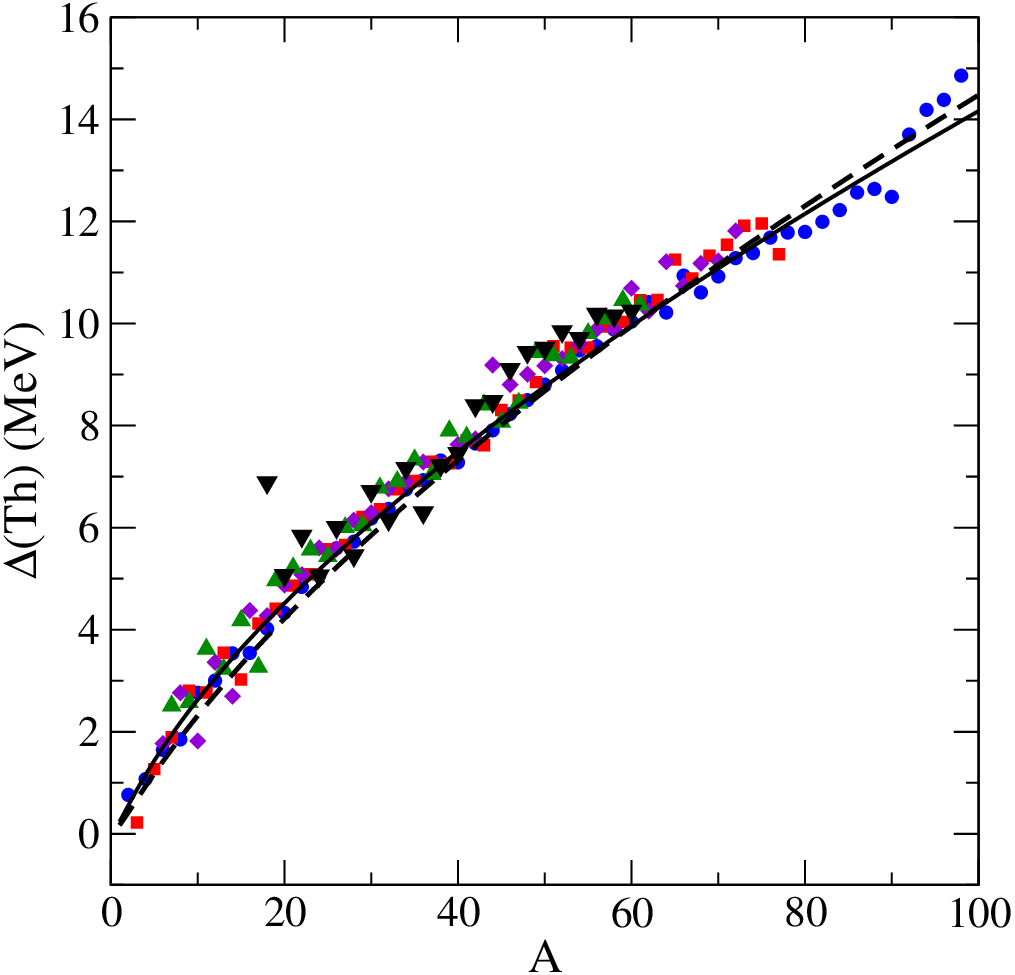}}
\caption{\label{Fullset} (color online)
The Coulomb shifts $\Delta(Th)$ 
plotted against the mass number ($A$) of the core nuclei
for isospins $T \le 2$. 
The isospin of the sets are displayed by
circles ($T=0$), squares ($T = 0.5$), diamonds ($T = 1$), up-triangles
($T = 1.5$), and down-triangles ($T = 2$).}  
\end{figure}

Using the Ame2003 mass tabulation, the excitation energies of nucleon 
emission thresholds in pairs of mirror systems were evaluated and the
results are displayed in Fig.~\ref{Thresholds} for all nuclei with known mass
($A$) and  with isospin ($T$) less than 3.  The excitation energies  
are mostly positive with an odd-even staggering with the (core) mass. 
The zero value is emphasised by the dashed lines in each panel. 
Negative values mean that the compound nucleus lies
beyond the appropriate nucleon drip line.  Mostly those are proton
emissive systems. 
 The non-zero isospin results have mirror core nuclei ($X, Y$ as denoted
above).  These excitations also show
odd-even mass staggering and,
with the $Th(pY)$ particularly, involve many more compound systems
lying beyond the proton drip line.
The associated $\Delta(Th)$ are plotted in Fig.~\ref{Fullset} in which
the curves are theoretical
results for $T = 0$ core nuclei ($N = Z = \frac{A}{2}$) with 
a proton. They were found from
\begin{equation}
\Delta(Th) =  \frac{\alpha  Z  \hbar c}{R}
= \frac{197.3269602}{137.035999679} \ Z\ \frac{1}{R} .
\label{deleq}
\end{equation}
where $R$ is as
recently defined~\cite{Di09} on adding an additional
proton radius, $r_p$, i.e.,
\begin{equation}
R = c_1\ A^{\frac{1}{3}} + c_2\ A^{-\frac{2}{3}}\  + r_p .
\label{radeq}
\end{equation} 
Ref.~\cite{Di09} specifies $c_1 = 0.94$ and \mbox{$c_2 = 2.81$~fm.}
Then by taking the proton radius to be
$r_p = 0.5$ fm, the result displayed by the dashed
curve in Fig.~\ref{Fullset} was found.
The second result displayed by the solid curve was found using a radius 
defined without any proton radius correction ($r_p = 0$) and 
making a nonlinear curve fit
to the T=0 data set to determine the coefficients. That resulted in
values of $c_1 = 1.07585$ fm and of $c_2 = 1.95514$ fm.   

\begin{figure}
\scalebox{0.45}{\includegraphics*{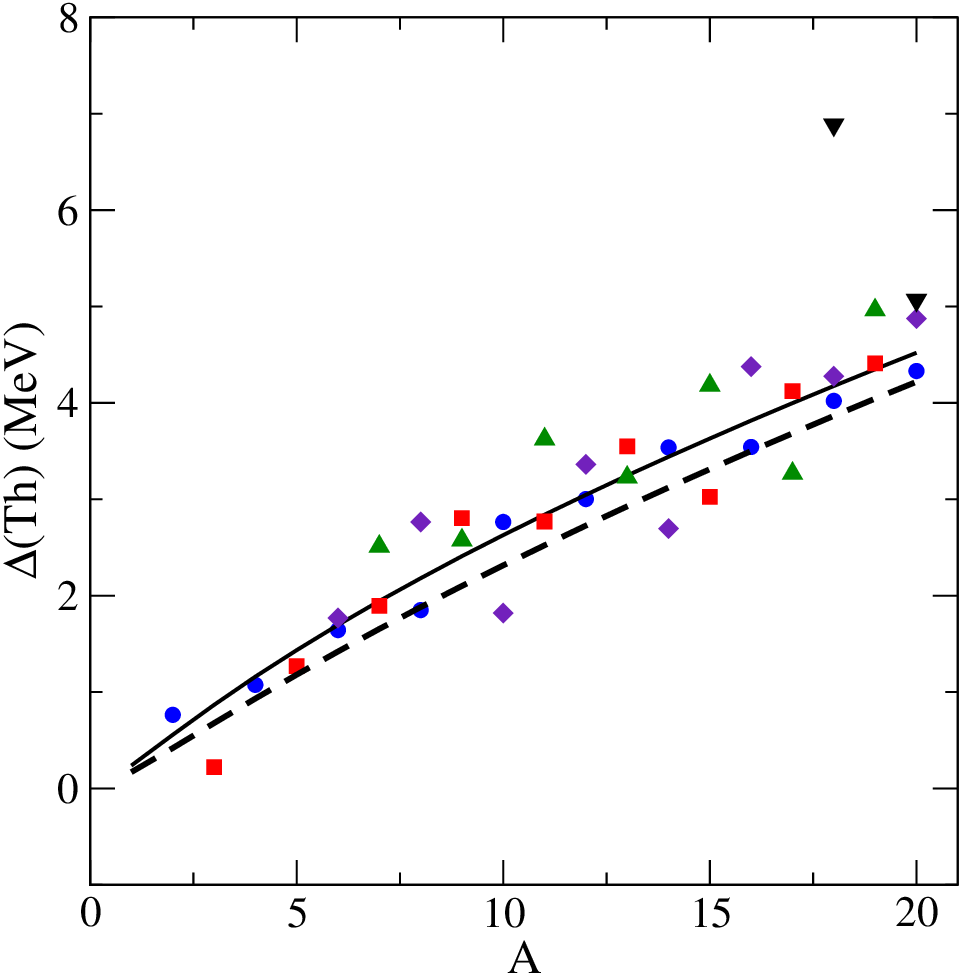}}
\caption{\label{Part-TH} (color online)
The Coulomb shifts $\Delta(Th)$ for the light mass
core nuclei. Notation is as used with Fig.~\ref{Fullset}}.
\end{figure}

The comparisons of the light mass results
($A \le 20$) are shown on larger scale in Fig~\ref{Part-TH}.
The two theoretical results are very good representations of the 
$T=0$ data set (filled circles) and they are good representations
of all the data save for one stand-out data point, associated 
with the mirror pair case of ${}^{19}$N$ \left({}^{18}\right.$N + 
$\left. n\right) - {}^{19}$Mg$\left({}^{18}\right.$Na +$\left.
p\right)$. Of these four nuclei, there is some information on ${}^{18,19}$N but
there is little or nothing known about the nuclei ${}^{19}$Mg and ${}^{18}$Na.
The latter two are  very exotic both being particle emissive.
With all other systems, the difference between excitation values of mirror 
system threshold energies show a gradual trend from $\sim 1$ to $\sim 5$ MeV
over the range of light masses (to $A = 20$).  
Though not universal, that difference tends to increase with isospin. 
The mirror nuclei
${}^{17}$C and ${}^{17}$Na are a mirror pair of current 
interest~\cite{Ti10,Fo10} and one might expect
the difference in the $n$+${}^{16}$C and $p$+${}^{16}$Ne threshold
energies relative to the ground states of ${}^{17}$C and ${}^{17}$Na
respectively, to be $\sim 4$ MeV.   Given that ${}^{17}$Na lies
beyond the proton drip-line, and that the $n$+${}^{16}$C threshold lies
below the ${}^{17}$C ground state by 0.728 MeV,  this suggests that 
the ground state  of ${}^{17}$Na would 
be $\sim 3.3$ MeV above the $p$+${}^{16}$Ne threshold; an MeV
larger than the recently anticipated~\cite{Ti10} value (2.4 MeV). 

Other mass formulae have been proposed in the past~\cite{Ke66,An85}. 
The first~\cite{Ke66} gave a formula for the mass excess
of nuclei from which a value of 35.61 MeV was suggested for ${}^{17}$Na.
Using the same formula to estimate the mass excess of ${}^{16}$Ne, they
suggest that the $p$+${}^{16}$Ne threshold would lie  3.65 MeV below,
i.e. $Th(p{}^{16}$Ne) = 3.65 MeV. 
 They also note the two proton-emission threshold would be 3.42 MeV 
below the ${}^{17}$Na ground state. In~\cite{An85} a
formula for isobaric mass multiplet energies for $A < 40$ was given, 
but only for
multiplets with $T \le 2$.  That isospin limit does not include the
sextuplet of nuclei of which ${}^{17}$C and ${}^{17}$Na are members.
However, in their Fig.~5 they did show  a plot of all isovector Coulomb
energies. That plot is very similar to the one for $\Delta(Th)$
given in Fig.~\ref{Fullset}. Their isovector Coulomb
energies for masses $A=17$ and $A=40$ are $\sim 4$ and $\sim 7$ MeV
respectively; values close to those we find for $\Delta(Th)$.

The mass of the exotic nucleus, $^{17}$Na, can also be estimated from
another energy systematics.
The energy differences of ground states of light mass 
isobars from one of
their most stable is well approximated by~\cite{An85}
\begin{align}
\epsilon =  \epsilon(Z,A) = & E(Z,A) - E(Z_s,A)
\nonumber\\ 
E(Z,A) & = M_{Z,A} - Z M_{\rm H} - (A-Z) M_{\rm n}
\nonumber\\ 
&\hspace*{2.0cm} 
- 0.6 Z(Z-1) A^{\frac{1}{3}} . 
\label{E-form}
\end{align}
Here, $Z_s$ is the charge number of the base nucleus, $M_{\rm n}$
is the mass of the neutron, 
$M_{\rm H}$ is the mass energy of the hydrogen atom.
The units are MeV and the last term is the approximation for
the Coulomb energy. 

We have used Eq.~(\ref{E-form}) to estimate
ground state energies of nuclei above one of the most stable of the isobars
for masses 6 to 20 to compare against the tabulations given in
the TUNL compilation~\cite{Ti93,Ti04}. Values of those estimates
are shown in Table~\ref{enucl}. 
\begin{table}[h]
\begin{ruledtabular}
\caption{\label{enucl}
Ground state gap energies (in MeV) of light mass isobars
determined using Eq.~(\ref{E-form}).}
\begin{tabular}{cccc|cccc}
 Base & A & Z & $\epsilon(Z,A)$ & Base & A & Z & $\epsilon(Z,A)$ \\ 
\hline
         & 6 & 1 & 28.19  &        & 7 & 2 & 11.66 \\
  $^6$Li & 6 & 2 & \ 4.05 & $^7$Li & 7 & 4 & $-$0.24 \\
         & 6 & 4 & \ 3.09 &        & 7 & 5 & 10.13  \\
\hline
         & 8 & 2 & 28.09 &         & 9 & 2 & 30.91 \\
  $^8$Be & 8 & 3 & 17.02 &         & 9 & 3 & 14.55 \\ 
         & 8 & 5 & 16.36 & $^9$Be  & 9 & 5 & $-$0.46 \\
         & 8 & 6 & 26.32 &         & 9 & 6 & 13.93 \\
\hline
         &10 & 2 & 39.42  &        &   &   &       \\ 
         &10 & 3 & 23.33  &        &11 & 3 & 34.34 \\
$^{10}$B &10 & 4 & \ 2.00 &        &11 & 4 & 12.88 \\
         &10 & 6 & \ 1.64 & $^{11}$B &11 & 6 & \ 0.07 \\
         &10 & 7 & 22.18 &         &11 & 7 & 11.26  \\
\hline
         &12 & 4 & 28.23 &         &   &   &        \\
$^{12}$C &12 & 5 & 15.21 &         &13 & 5 & 15.21 \\
         &12 & 7 & 14.97 & $^{13}$C &13 & 7 & $-$0.06 \\
         &12 & 8 & 26.80 &         &13 & 8 & 14.92 \\
\hline   
         &14 & 5 & 24.72 &         &15 & 5 & 32.66 \\
$^{14}$N &14 & 6 & \ 2.36 &        &15 & 6 & 11.91 \\
         &14 & 8 & \ 2.40 & $^{15}$N &15 & 8 &  0.13 \\
         &   &   &        &        &15 & 9 & 10.94 \\
\hline   
         &16 & 6 & 23.06 &         &17 & 6 & 26.35 \\  
$^{16}$O &16 & 7 & 12.97 &         &17 & 7 & 11.16 \\
         &16 & 9 & 12.39 & $^{17}$O &17 & 9 & $-$0.19 \\   
         &16 &10 & 22.20 &         &17 &10 & 10.90 \\
\hline      
         &18 & 6 & 31.32 &         &19 & 6 & 41.00 \\ 
         &18 & 7 & 17.54 &         &19 & 7 & 22.53 \\
$^{18}$F &18 & 8 & \ 1.23 &        &19 & 8 & \ 7.64 \\  
         &18 &10 & \ 1.10 & $^{19}$F &19 &10 & $-$0.03 \\
         &   &   &        &        &19 &11 & \ 7.43  
\end{tabular}
\end{ruledtabular}
\end{table}
Therein even and odd mass results are shown in the left
and right  most set of columns respectively. The nucleus
identified in the columns headed by `base' are those used
with the charge numbers $Z_s$ in each group. With the even-mass sets, the base entry lies on the line containing an energy
that will be compared with one immediately below in what follows.
The base energy for the odd-mass sets lies on the line about which
pairs of results will be compared.
Almost all of these results agree with the listed values~\cite{Ti93,Ti04}
to better than a percent.  Those that do not are
${}^7$B (10.44 vs 10.13), ${}^{11}$N (12.13 vs 11.26), and 
${}^{13}$N (+0.06 vs -0.06). There is a close pairing of nuclei
according to their isobaric spin, as has been noted before~\cite{An85}. 

\begin{figure}[h]
\scalebox{0.45}{\includegraphics*{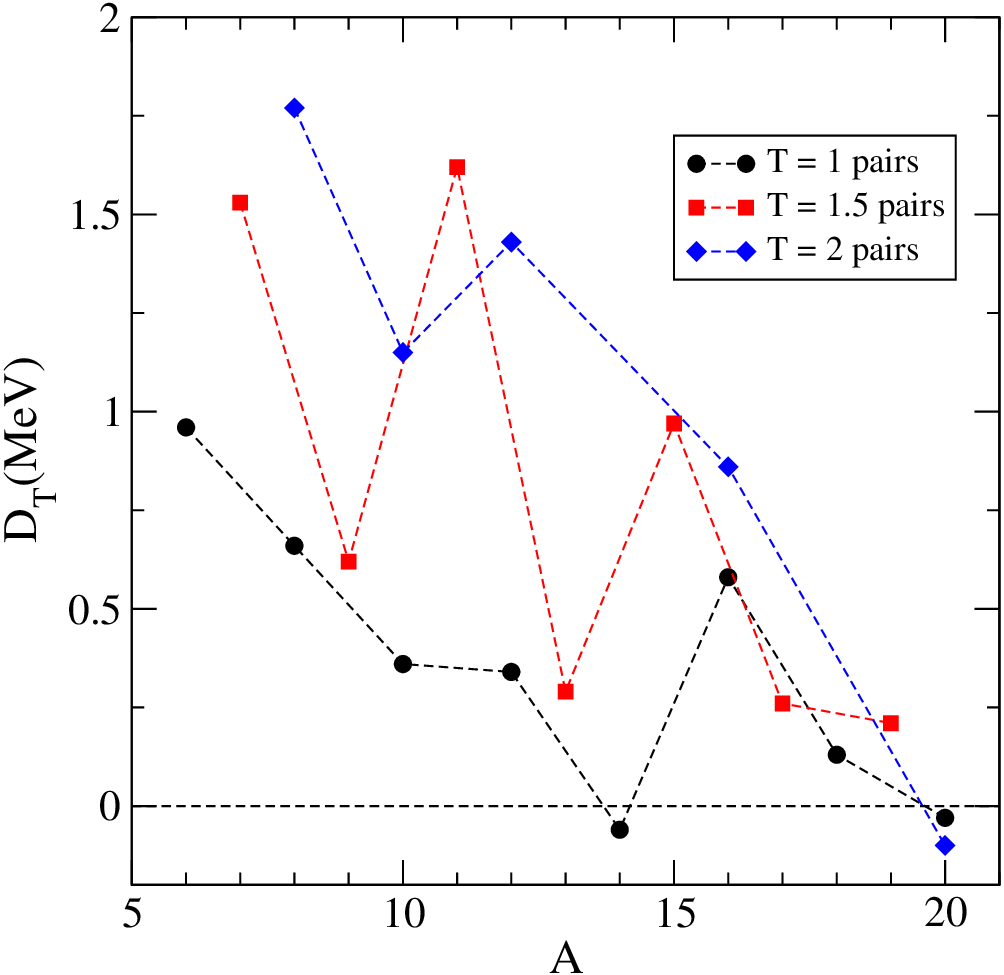}}
\caption{\label{IsoE} (color online)
The energy differences between mirror ground state energies calculated using
Eq.~(\ref{E-form}).
}
\end{figure}
For these light-mass nuclei, the energy differences
between ground states of
isotopic spin pairs, 
$D_T (= \epsilon(Z,A) - \epsilon(A-Z,A))$,
are plotted in Fig.~\ref{IsoE}. 
Of note is that for the three separate isospin values, the trend is for
the energy differences of the pairs to decrease as mass increases. 
We surmise that
this will be the case for pairs with higher isospin values, and for
the $T=$ 2.5 pair of ${}^{17}$C and ${}^{17}$Na of particular interest
herein.  
With that surmise,
$\epsilon(11,17)$ (for ${}^{17}$Na) would be $\sim 25.5 \pm 0.5$ MeV above the 
${}^{17}$O ground state.

By inverting Eq.~(\ref{E-form}), we can find an atomic
mass-energy for exotic nuclei, which for mass-17 systems
are listed in Table~\ref{masse}.
The values of $\epsilon(Z,A)$ for the five known nuclei were
taken from the TUNL tabulation~\cite{Ti93,Ti04} and the atomic
mass-energies for the neutron, ${}^1$H, and all mass-16
nuclei required were taken from the mass tabulation of
Ame2003~\cite{Au03}. 
All evaluated masses of the known mass-17 nuclei agree
with the tabulated ones~\cite{Au03} to better than 1 part in a million
so that the derived nucleon-core nucleus thresholds ($Th(nX)$ and  $Th(pY)$)
also agree with tabulated values~\cite{Ti93,Ti04}.

\begin{table}[h]
\begin{ruledtabular}
\caption{\label{masse} Mass-17 system properties deduced from 
inversion of Eq.~(\ref{E-form}). The base system defining $E(Z_s,A)$
is ${}^{17}$O. All energies are in units of MeV and
masses are in units of $u$.}
\begin{tabular}{ccccc}
Nucleus & $\epsilon(Z,A)$ & $M_{Z,A}$ & $Th(nX)$ & $Th(pY)$\\
\hline
 ${}^{17}$C  & 26.35 & 17.02259 & 0.73 & 23.33 \\
 ${}^{17}$N  & 11.16 & 17.00845 & 5.88 & 13.11 \\
 ${}^{17}$O  & 0.0   & 16.99914 & 4.14 & 13.78 \\
 ${}^{17}$F  &$-$0.19& 17.00209 & 16.8 &\  0.60 \\
 ${}^{17}$Ne & 10.92 & 17.01778 & 15.6 &\  1.49 \\
\hline
 ${}^{17}$Na & 25.5  & 17.03752 & 26.8 & $-$3.66 \\
\end{tabular}
\end{ruledtabular}
\end{table}

Consequently, assuming a gap energy for ${}^{17}$Na 
above the base line of ${}^{17}$O  to be 25.5$\pm$ 0.5,
the atomic mass-energy of $M_{^{17}Na}$ would be 17.03752 $\pm$ 
0.00054 u; 3.66 $\pm$ 0.5 MeV above the 
proton-${}^{16}$Ne threshold.

In summary,
structure models ~\cite{Ti10,Fo10} have suggested that the centroid of the
resonant ground state of $^{17}$Na relative to
the $p+^{16}$Ne threshold is 2.4 and 2.7 MeV. A previous mass formula 
suggested a value of 3.65 MeV. 
Considering available particle emission
data, and a theoretical formulation of the same, we assess it to be
$~3.3$ MeV, and from trends in differences of ground state energies in
light mass isobars, we estimate 3.66 $\pm$ 0.5 MeV. 


\bibliography{thresh}

\end{document}